\documentclass[prd,floats,preprint,superscriptaddress,eqsecnum,floatfix,nofootinbib,12pt]{revtex4}
\pdfoutput=1
\usepackage{amsmath}
\usepackage{graphics, graphicx}
\usepackage{epsfig}
\usepackage{ulem}
\usepackage{color}

\def\1p{{(1p)}}
\def\pfp{p^{(1p)}}
\def\be{\begin{equation}}
\def\ee{\end{equation}}
\def\beq{\begin{eqnarray}}
\def\eeq{\end{eqnarray}}
\def\p0{\phi_0}
\def\z0{\zeta_0}

\def\aoi{D^{\ge 1}}

\def\Dloc{D_{\rm loc}}

\def\phiei{\phi_{\rm ei}}

\def\3G{^3{\cal G}}

\def\ol2{\frac{1}{\ell^2}}

\def\Dul0{D_{\ell}(t_0)}
\def\Dl{D_{\rm loc}}

\def\red{red}
\def\blue{blue}
\def\aoe{E^\ge}
\def\HP{$(I,\Psi) \ $}
\def\jf{}
\def\rf{}
\def\hf{}
\def\lf{}
\def\mf{}

\newcommand{\ttle}[1]{{\it #1}}

\begin{document}

\vspace{1cm}

\title{The Observer Strikes Back\footnote{Based on a talk given by JH at the conference on the Philosophy of Cosmology, Tenerife, Spain, September 13-15, 2014.}}

\author{James Hartle}
\affiliation{Department of Physics, University of California, Santa Barbara,  93106, USA}
\author{Thomas Hertog}
\affiliation{Institute for Theoretical Physics, KU Leuven, 3001 Leuven, Belgium}

\bibliographystyle{unsrt}


\begin{abstract}
In the modern quantum mechanics of cosmology observers are physical systems within the universe. They have no preferred role in the formulation of the theory nor in its predictions of third person probabilities of what occurs. However, observers return to importance for the prediction of first person probabilities for what we observe of the universe: What is most probable to be observed is not necessarily what is most probable to occur. This essay reviews the basic framework for the computation of first person probabilities in quantum cosmology starting with an analysis of very simple models. It is shown that anthropic selection is automatic in this framework, because there is no probability for us to observe what is where we cannot exist. First person probabilities generally favor larger universes resulting from inflation where there are more places for us to be. In very large universes it is probable that our observational situation is duplicated elsewhere. The calculation of first person probabilities then requires a specification of whether our particular situation is assumed to be typical of all the others. It is the combination of the model of the observational situation, including this typicality assumption, and the third person theory which is tested by observation. We conclude with a discussion of the first person predictions of cosmological observables such as the cosmological constant and features of the primordial density fluctuations, in the no-boundary quantum state of the universe and a dynamical theory in which these are allowed to vary.

\end{abstract}

\vskip.8in
\vspace{1cm}


\maketitle


\section{Introduction}
\label{intro}

The context for this essay is our universe --- the whole closed system at all times containing all the galaxies, stars, planets, biota, human societies, you, us, etc.  There is nothing outside. 
Two kinds of description of the universe can be distinguished:

{\it Third person Descriptions:} Descriptions of what the universe contains and how that evolves ---  histories of what occurs. 

{\it First Person Descriptions:} Descriptions of what we as the collection of human scientists observe of the universe {\rf and use to test cosmological models}.  

\noindent The connection between these two kinds of description is the subject of this essay. 

Quantum mechanical theories, and also classical ones, provide probabilities for different descriptions. Correspondingly we can distinguish two different kinds of probabilities for any observable $\cal O$.
Third person probabilities\footnote{Following \cite{HH06}, in most of our previous work we have used `bottom-up probabilities' and `top-down probabilities' for what are called here `third-person probabilities' and `first person probabilities'.}
 
\begin{subequations}
\be 
p(\cal O)
\ee
for what {\rf values of} $\cal O$ occur, and first person probabilities 
 \be
 p^\1p(\cal O)
 \ee
 \end{subequations}
for what {\rf values of} $\cal O$ we observe. The connection between these two kinds of probabilities is the focus of this paper. 

We will consider theories whose direct outputs are third person probabilities for histories of what occurs in the universe. These include probabilities for the existence, evolution, and {\rf functioning } of any observing subsystems such as ourselves. Such subsystems play  no special role in formulating the theory --- they are just one kind of subsystem among many. We refer to such theories as `third person' theories. {\rf Our most successful theories of cosmology are of this kind.} Classical physics {\rf in general} is an example of a third person theory. In quantum mechanics the extensions of the ideas of Everett \cite{MW} are third person theories, including those used in this paper.

Observers do play a preferred role in calculating first person probabilities for observations from third person ones.  First person probabilities are third person ones conditioned on a description of the observational situation. We shall use a variety of specific models to infer the following general conclusions:
\begin{itemize}

\item What is most probable to occur is not necessarily what is most probable to be observed. 

\item Anthropic selection is an automatic consequence of first person probabilities.

\item In universes large enough that we may be duplicated as physical subsystems elsewhere, {\rf the description of the observational situation needed to compute first person probabilities must also specify} whether our particular situation is typical of all the others. It is the combination of {\rf the model of the observer, including this typicality assumption,} and the third person theory which is tested by observation. 

\end{itemize}

Observers and their observations are of central importance in the formulation of Copenhagen quantum mechanics. In the quantum mechanics of closed systems observers might seem to have been demoted to the status of one subsystem among many. Indeed, they have little effect on third person probabilities. But, as a consequences of the conclusions above,  the observer returns to importance in the calculation of first person probabilities for observations by which the theory is used and tested. The observer strikes back. 

Section \ref{theory} {\rf sketches the framework} of the third person theory we will employ. Section \ref{Hvol} discusses issues involved with first person probabilities. Section \ref{boxmodels} uses a model universe to make more concrete the notions of first and third person descriptions of the universe and their associated probabilities. Here we also describe the connection between third and first person probabilities in a set of simple models. Section \ref{anthropic} describes how anthropic selection {\rf emerges automatically as a feature of a certain class of} first person probabilities. Section \ref{cgrain} shows how first person probabilities can sometimes be calculated directly from the theory with an appropriate coarse graining. {\rf Section \ref{examples} discusses the first person predictions of a number of cosmological observables, such as the cosmological constant, in cosmological models based on the the no-boundary quantum state of the universe and a dynamical landscape theory in which these observables can take a range of values.} In Section \ref{conclusion} we try to set our results in a more general view of physical theories.


\section{Third Person Quantum Cosmology}
\label{theory}

This section briefly describes the elements of the theory that predicts third person probabilities for histories of the universe {\hf which we use}  to reach the general conclusions mentioned in the introduction\footnote{For more details the reader can consult the authors' papers on which this essay is implicitly based and through them find further references. For the quantum mechanics of closed systems see e.g. \cite{Har93a,Har95c}. For quantum cosmology see e.g. \cite{HH06,HHH08a,HHH08b,HH09,HHH10,HHH10b}. There is a little more detail about the no-boundary quantum state of the universe \cite{HH83} in Appendix \ref{NBWF-mini}.}

{\rf We view} the universe as a closed quantum mechanical system. It contains everything. Galaxies, stars, planets, their biota, observers (including us!) etc are physical subsystems of the universe subject to its quantum mechanical laws consistent with an Everettian point of view \cite{MW}. The basic variables describing the universe and its contents are four-dimensional cosmological spacetime geometries and four-dimensional configurations of matter fields.  The basic ingredients of the theory are an action $I$ describing the dynamics of geometry coupled to matter fields, and a quantum state of the universe $\Psi$. We denote the theory as \HP. 

The theory \HP predicts third person probabilities for the individual members of sets of alternative {\rf four-dimensional} histories of the universe, including those histories that describe its classical evolution. In this way \HP can supply third person probabilities for such large scale features of the universe as the amount of inflation, the approximate homogeneity and isotropy, {\rf the pattern of cosmic microwave background (CMB) variations}, and the formation and evolution of the distribution of galaxies. In principle \HP also supplies third person probabilities for the accidents of biological evolution, the existence of observers like ourselves, etc that are well beyond our power to compute or even estimate. For the examples in this paper we will mostly use {\rf histories in which we }are at a single moment of time. The time approximately 14 Gyr after the big bang when our observations of the universe are made. 
The theory \HP is an example of what we will call a third person theory.

We test and utilize a theory not by its third person probabilities for what occurs, but by its first person probabilities for what we observe. 
{\rf To compute first person probabilities we first need to model the observational situation which we do next.}

\section{First Person Quantum Cosmology}
\label{Hvol}

We begin by recalling the definition of a Hubble volume.  We cannot see further in the universe than the distance that light travels to us from the big bang, roughly 14Gyr ago. A present volume of the same size as this distance is called a `Hubble volume'. In order of magnitude the distance is $c/H_0$ where $H_0$ is the present Hubble constant. This size is approximately $4000$ Megaparsec or $10^{23 }{\rm \ km}$. This is the largest scale over which we can currently observe the universe.

As observers we are physical systems within the universe with only a probability to have evolved in any one Hubble volume and a probability to be replicated in other Hubble volumes if the universe has a very large number of them. An observer is a very special kind of fluctuation $D_0$ in the universe. It's a fluctuation that is not singled out by quantum theory from, say, density fluctuations that produced the CMB.  But the probability for observers is very difficult to compute. We will therefore employ a highly simplified model of observers: All observers are alike (copies of us) and either exist in any Hubble volume with a third person probability $p_E (D_0)$ or do not exist with a probability $1-p_E$.  Realistically the probability $p_E$ incorporates the probability of the accidents of several billion years of biological evolution. Therefore, whatever its value is, it is very, very, very small. This is a very crude model of complex observers, but better than many treatments where the probability that observing systems evolved {\rf as part of the universe's evolution} is not considered at all.

{\rf Since both we and what we observe are part of the universe, first person probabilities can be computed from the third person ones. If we are unique as physical systems within the universe the first person probabilities are {\rf simply} third person ones for what's observed conditioned on a description of the observational situation like the one given above, in terms of probabilities $p_E$ for data $D_0$.
But if we are not unique then {\rf a more careful specification of} the observer is required which includes an assumption about which instance of $D_0$ is us, or more generally a probability distribution on the set of copies {\hf called a xerographic distribution} \cite{SH10}. In this essay we will make the minimal assumption that we are equally likely to be any of the incidences of $D_0$ that the third person theory \HP predicts. {\hf The way to make other assumptions is discussed in Appendix \ref{typicality}.}

It's a common intuition that the presence or absence of observers is unimportant for the behavior of the universe on cosmological scales because observers are generally small subsystems. That intuition is correct for third person probabilities, but it is not correct for first person ones. As physical systems we have only a tiny probability to exist in any one Hubble volume. That is why third person probabilities are little affected by the presence or absence of us. But this also means that we have a greater probability to live in a larger universe than a smaller one, because in the larger there are more Hubble volumes in which to be. Therefore, even if the third person probabilities favor smaller universes the first person ones may favor larger ones. As we will show by example, what is most probable to occur (third person) is not necessarily what is most probable to be observed (first person).  It is in this way that the observer returns to importance in cosmology.

\section{Toy Model Universes}
\label{boxmodels}

This section derives the connection between third and first person probabilities for a very simple class of models that provide  elementary examples showing what is most probable to be observed is not necessarily what is most probable to occur.
In the trade these are called `box models' \cite{HS07}. 

\subsection{Third person description and probabilities}
\label{box3}

Histories of the universe at {\hf one time are}  modeled as a collection of $N$ boxes representing Hubble volumes at that time. Each box has a color --- say red or blue --- modeling different CMB maps. Each box may or may not contain an observer.   A third person description of a history specifies the number of boxes, their color, and whether each box contains an observer or does not. {\hf A third person quantum theory specifies probabilities} for these histories --- a probability for the number of boxes, a probability for the color of each box, and the probability $p_E$  for whether an observer exists in each. 
\begin{figure}[t]
\includegraphics[width=6in]{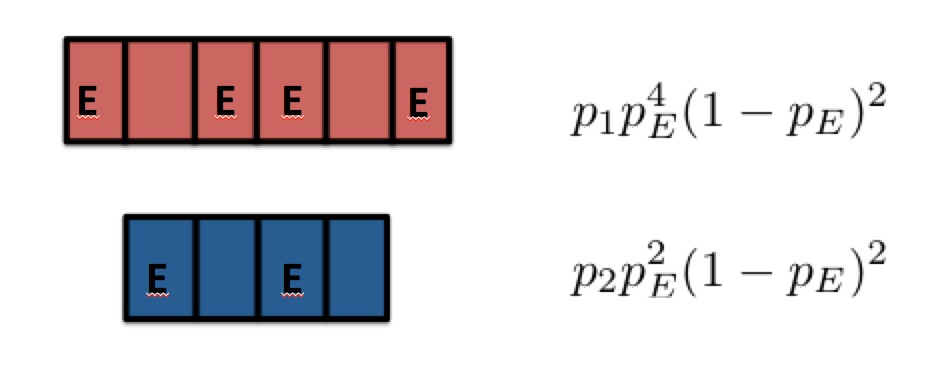}\hfill
\caption{Two histories of the simple box model universe described in the text. The boxes model Hubble volumes. Their color models an observable like the CMB. An `E' means that an observer is in the box observing its color. A blank means there is no observer in the box. The third person probabilities for these histories to occur are at the right.}
\label{box1}
\end{figure}

We illustrate this with the very simple set of single time histories \cite{HS07} shown in Figure \ref{box1}. At one time, only two possible histories of $N$ and color  are possible--- $N_1$ boxes all \red \ that occurs with a third person probability $p_1$, and $N_2$ boxes all \blue \ that occurs with third person probability $p_2=1-p_1$. The probability that there is an observer in any box is $p_E$ --- the same for all boxes. The probability that there is no observer in a box is $1-p_E$. The third person probability of a history {\rf with a specific set of} $n_E$ of $N_k$ boxes occupied by observers is
\be
\label{pboxmodel}
p_k p_{E}^{n_E} (1-p_E)^{N_k-n_E}
\ee
where $k=1 \  \text{(all red)}$ or $k=2 \  \text{(all blue)}$. The complete set of histories consists of all red and all blue histories, with different numbers of boxes in each, and the various possible ways the boxes can be occupied by observers.

\subsection{First Person Probabilities for Observation}
\label{1stbox}

We are one of the observers in one of the histories. We now ask for the theory's prediction for the first person probability that we observe red (WOR). To calculate that, assume that in  either history we are equally likely  to be any one of the occurrences of $E$.  Then the probability that we observe red (WOR) is evidently the probability that we are in the history with all red boxes ($k=1$). 

The probability that WOR is {\it not} the probability $p_1$ that the history with all red boxes occurs because that could happen with no boxes occupied by observers. Rather the probability for WOR is the probability that history $k=1$ occurs with {\it at least one observer}. Denoting `at least one $E$' by $E^\ge$ we have
\begin{align}
\label{atleastone}
p(\aoe \ \text{in 1}) =& \ 1 - p(\rm{no} \ E \  \text{in 1}) \nonumber \\
               =&\  1 - (1-p_E)^{N_1}  .
 \end{align}
 \begin{subequations}
 \label{obsprob}
 The normalized probability that we observe red (WOR) is then
 \be
 \label{WOR}
 \pfp(WOR) = \frac{p_1 [ 1-(1-p_E)^{N_1}]}{\sum_k p_k [ 1-(1-p_E)^{N_k}]} \  .
 \ee
 Similarly the probability that we observe blue is 
  \be
 \label{WOB}
 \pfp(WOB) = \frac{p_2 [ 1-(1-p_E)^{N_2}]}{\sum_k p_k [ 1-(1-p_E)^{N_k}]} \  .
 \ee
 \end{subequations}
We now discuss important limiting cases. 

{\it Rare in all histories:}  When both $p_E N_1\ll 1$ and $p_E N_2 \ll 1$ physical systems like us occur only rarely in each of the two possible histories. {\jf Therefore we can assume that as a physical subsystem we are unique in the universe.}  The probabilities for color observation \eqref{obsprob} then become
\begin{subequations}
\label{volwgt}
\begin{align}
\label{rare}
\pfp(WOR) &\approx \frac{N_1 p_1}{N_1p_1+N_2p_2}, \\ 
\pfp(WOB) &\approx \frac{N_2 p_2}{N_1p_1+N_2p_2}  \ .
\end{align}
\end{subequations}
The first person probabilities for our observation of red or blue are the third person probabilities of the all red and all blue histories {\it weighted} by the number of Hubble volumes in each.  This is called `volume weighting'. It favors larger universes where there are more places for us to occur as has been extensively discussed in cosmology (e.g. 
\cite{Page97,Hawking07,HHH08a}).

It's important to emphasize that volume weighting is not an extra assumption in addition to the theory of the histories. Rather it is a straightforward consequence of that theory {\rf in models where we are rare in all histories.}

In a third person description of this model blue is more likely to occur if $p_2>p_1$. But red is more likely to be observed if $N_1p_1>N_2p_2$. Thus we have an elementary example of what is most probable to be observed is not necessarily what is most likely to occur. That is the return of the observer.

Other limiting cases are also interesting:

 {\it Common in both histories:}  When both $p_E N_1\gg 1$ and $p_E N_2 \gg 1$, copies of us as physical systems are common in both histories. Then the probabilities \eqref{obsprob} are 
 \begin{subequations}
\label{common}
\begin{align}
\label{rare2}
\pfp(WOR) &\approx p_1, \\ 
\pfp(WOB) &\approx p_2 . 
\end{align}
\end{subequations}
Thus when all histories in the ensemble are very large universes what is predicted to be observed is also what is predicted to occur. 

 {\it Rare in one history, common in the other:} When, say, $p_E N_1\gg 1$ but $p_E N_2\ll 1$ copies of us as physical systems are common in the all red history and rare in the blue one. We have $[1-(1-p_E)^{N_1}] \approx 1$ and $[1-(1-p_E)^{N_2}] \approx N_2p_E \approx 0$ since $p_E$ is very, very small.
 Then, 
\begin{subequations}
\label{both}
\begin{align}
\label{rare1}
\pfp(WOR) &\approx 1, \\ 
\pfp(WOB) &\approx 0.
\end{align}
\end{subequations}
 This is the case where the return of the observer has its most striking effect. The first person probabilities select large histories over small ones {\rf even when the latter have larger third person probabilities}. We will see a concrete example of this in a more realistic model in Section \ref{EI}.

\subsection{A Crisis of Computability?}   
\label{computability}
Box models divide the theory for predicting first person probabilities for what we observe into two parts. First, there is the specification of the third person probabilities $p_k$  for the large scale features of the models --- the number of Hubble volumes and the color of each. Second there are the third person probabilities for the occurrence of observers inside each Hubble volume, summarized by the one number $p_E$, {\rf which are used to describe and to condition on the observational situation.}

As we will see in Section \ref{examples}, it  is possible to make computationally tractable calculations of $p_k$ in simple models. 
But the probability $p_E$ would naturally include the probability that human observers evolved in a Hubble volume. To calculate this, or even estimate it, would involve considering several billion years of the chance accidents of biological evolution. This is well beyond our power to compute even assuming that we have a theory that is well enough formulated to define the task. 

It is therefore fortunate that in all of the interesting limiting cases discussed above the probability $p_E$ cancels out. Thus, it is in the regime of universes so small that we are unique, or universes so large that we are common,  that observations are easily and objectively calculated.

\section{Anthropic Selection is Automatic in Quantum Cosmology}
\label{anthropic}

Consider again the simple box model in Section \ref{1stbox} but suppose that the probability for an observer to occupy a red box $p_E^R$ is different from the probability $p_E^B$ to occupy a blue box. Suppose further that for some reason red is necessary for observers so that the probability $p_E^B$ to occupy a blue box is exceedingly small. It is then obvious, and easily worked out, that the first person probability that we observe red is very near unity. This is a very simple example of anthropic selection: The all red history is selected because observers do not exist in the alternative.

The key point here is that anthropic selection is automatic. No additional assumptions or principles were needed beyond the probabilities $p_1$, $p_2$, $p_E^R$ and $p_E^B$ which are all third person probabilities following from the underlying theory. {\rf Anthropic selection emerges as an intrinsic feature of} the first person probabilities for the observer's observations \cite{HH13}. 

{\rf This is the case also in more general and more realistic cosmological models. As an example} consider an ensemble of single time histories which we take to be at the present age of the universe $t_0 \approx 14 {\rm Gyr}$. Assume that the theory \HP predicts third person probabilities for what we may call a set of background histories each with the same number of Hubble volumes but differing in the value of the cosmological constant $\Lambda$, which is assumed positive and the same in all Hubble volumes of one history. The theory thus allows the cosmological constant to vary. The histories can be labeled by the value of $\Lambda$ and their third person probabilities written $p(\Lambda)$.  As a fluctuation on these backgrounds the theory \HP also predicts a probability that in any Hubble volume there occur data $D_0$ that describe our observational situation (but not including any record we might have of the value of $\Lambda$). These probabilities depend on $\Lambda$. The complete set of histories is thus labeled by the backgrounds and which Hubble volumes in them are occupied by $D_0$. 

Assume that we are typical of the incidences of $D_0$ in any one history.  The first person probability that we observe a value of $\Lambda$ (WO$\Lambda$) is the third person probability for the history $\Lambda$ conditioned on the existence of at least one instance of $D_0$ [cf. \eqref{atleastone} et seq.]. Using the Bayes identity we have  
\be
\label{bayes}
\pfp(WO\Lambda) \equiv p(\Lambda|\aoi_0) = \frac{p(\aoi_0|\Lambda) p(\Lambda)}{\sum_\Lambda p(\aoi_0|\Lambda) p(\Lambda)} . 
\ee
For values of $\Lambda$ for which  $p(\aoi_0|\Lambda)$ is negligibly small the probability that we observe that value will also be negligibly small. Hence anthropic selection of values of $\Lambda$ that are consistent with observers is automatic. This kind of argument assumes that the third person theory allows $\Lambda$ to vary over an anthropically allowed range  --- a range consistent with the evolution of $D_0$.  We give an example of such a model in Section \ref{EI}.  If however the theory determines a unique value of $\Lambda$, then either that must be consistent with $\aoi_0$ or the theory is incorrect (e.g. \cite{Har05}).

Barrow and Tipler \cite{Barrow86}, and Weinberg \cite{Wei89} have argued that the observed value of the cosmological constant could not be much larger than $\Lambda \sim 10^{-122}$ in Planck units, not far from its observed value. Were $\Lambda$ larger the universe would expand too rapidly for galaxies to have formed by the present age $t_0\sim 14 \ {\rm Gyr}$ and human observers would not be here. In our scheme the probability $p(\aoi_0|\Lambda)$ would be near zero. 

{\rf This kind of argument is an example of what one could call traditional anthropic reasoning. In this}, the anthropically allowed range of values of an observable like $\Lambda$ is determined from classical arguments like those involving galaxy formation mentioned above. It is then assumed that there is some unknown mechanism for $\Lambda$ to vary over this range. In the absence of a specific model of this mechanism a uniform distribution over the range is often assumed and the most probable $\Lambda$ {\rf predicted on the basis of purely anthropic arguments}. Impressively detailed calculations were carried out in this way {\rf with mixed results}, e.g \cite{Tegetal06}. 
But this program suffers from several uncertainties. For example, different results are obtained if different combinations of constants were assumed to vary \cite{LR05}. {\rf In addition, traditional anthropic reasoning is not part of any theoretical framework. Rather anthropic selection arises from an additional assumption or `principle'.} 

{\rf The chief difference with anthropic selection in quantum cosmology is that 1) there the theory \HP provides a mechanism for what constants vary and how they vary, 2) the observer is a physical subsystem with a certain probability predicted by the theory to evolve in any Hubble volume and 3) anthropic selection emerges automatically as a property of the first person probabilities by which we use and test the theory. To summarize, anthropic selection in quantum cosmology:}

\begin{itemize} 

\item Is not a choice.

\item Does not require invoking some `anthropic principle'. 

\item Does not change the objective nature of the underlying third person theory. 

\item Does require a typicality assumption if there is a chance we are replicated. 

\end{itemize}

The observer is important for first person probabilities for observations through anthropic selection.
In Section \ref{anth-L} we will describe a more precise calculation of $\pfp(WO\Lambda)$ {\rf in which the probabilities $p(\Lambda)$ in \eqref{bayes} are obtained from a concrete model of the universe's quantum state $\Psi$ in combination with a dynamical theory $I$ in which $\Lambda$ can vary.}

\section{A Remark on Coarse Graining}
\label{cgrain}
Our observations of the universe extend at most over a Hubble volume.  But  this may only be a tiny region in a vastly larger inflationary universes of the kind contemplated in contemporary  cosmology. Indeed, some calculations \cite{CREM} suggest that the universe {\rf typically} becomes spatially infinite as a consequence of eternal inflation. Much of the third-person information about what occurs in the universe on very large scales is irrelevant for the first person predictions for our observations in our Hubble volume, and perhaps not even well defined. In this section we illustrate how, in certain models, first person probabilities for our observations can be calculated directly using coarse grainings that ignore most of the structure outside our Hubble volume. {\hf Coarse graining is not an ad hoc assumption. It  is central and inevitable  in quantum mechanics, statistical physics, complexity, and many other areas of science (e.g. \cite{ GH07}).}

\begin{figure}[t]
\includegraphics[width=6in]{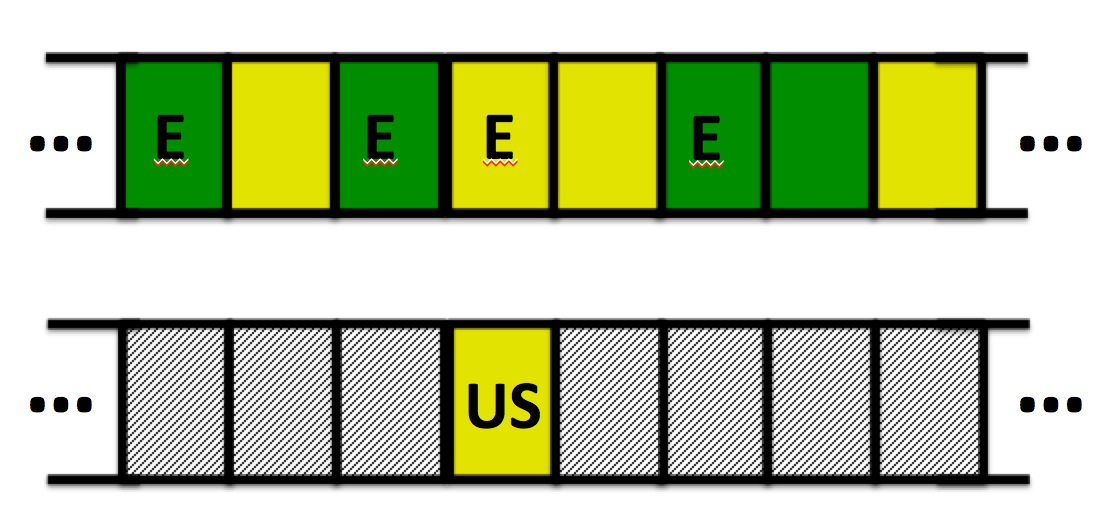}\hfill 
\caption{Fine and coarse-grained histories of  a box model discussed in the text. The boxes model Hubble volumes. Their color models an observable like the CMB in this case either yellow or green. An `E' means that an observer is in the box observing its color. The top history is fine grained with a color and E or not E in every box. The bottom history is coarse grained. The possibilities have been summed over for every box except one ---- our box. That enables a straightforward calculation of the first person probability that we see one color or the other. The details outside our box are irrelevant for this. }
\label{YGfig}
\end{figure}


We illustrate how {\hf coarse graining}  works with a simple box model of the kind used in Section \ref{boxmodels}. Consider a universe with an infinite set of boxes as illustrated in Figure \ref{YGfig}. Each box has a color. This is either yellow with probability $p_Y$ or green with probability $p_G=1-p_Y$. There is a probability $p_E$ for an observer to be in any box observing its color. Thus all the boxes are statistically identical. In this sense this universe has a discrete translation symmetry.

Now we ask for the first person probability that we observe yellow. The answer is obvious. We are in one box. All the boxes are statistically the same. The first person probability $\pfp(Y)$ for observing yellow is the same as the probability that any of the boxes is yellow, viz.  
\be
\label{obsyellow}
\pfp(Y)=p_Y.
\ee
Although obvious, it is instructive to see how this result follows from our general framework through coarse graining starting from third person probabilities. A fine grained history would specify the  color $(Y,G)$ of each box, and whether there exists an observer in it or not $(E, \bar E)$ (top in Figure \ref{YGfig}). The third person  probability for {\hf one particular}  history having $n_Y$ yellow boxes, $n_G$ green boxes,  $n_E$ boxes with observers, and $n_{\bar E}$ without,  is
\be
\label{fgprobs}
(p_Y)^{n_Y}(1-p_Y)^{n_G}(p_E)^{n_E}(1-p_E)^{n_{\bar E}} .
\ee
But these probabilities tend to zero in a universe with an infinite number of boxes. The probabilities for these fine-grained histories are not well defined. 

Finite probabilities can be obtained by coarse graining. {\hf That is, they can be obtained by summing \eqref{fgprobs} over what's irrelevant for our observations.} To get first person probabilities for our observations we can coarse grain over the alternatives in every box but ours. That means sum the probabilities \eqref{fgprobs} over the alternatives $(Y,G)$ and $(E,\bar E)$ giving a factor of unity for every box but ours (bottom in Figure \ref{YGfig}). That gives the first person probabilities.

This result generalizes straightforwardly to any finite number of kinds (colors) of Hubble volumes, {\rf to models where the probabilities $p_E$ depend on the color}, and to more than one configuration {\rf (history) }of boxes with third person probabilities like the $p(k)$ of Section \ref{box3}. {\rf In Section \ref{EI} we review an example of coarse graining used in realistic cosmology (see also \cite{HHH10b}).}

To summarize, in infinite or just very large universes {\hf focusing on our observations in our Hubble volume}  motivates coarse grainings that directly lead to well defined first person probabilities for observations. 
{\lf It is an intriguing open question whether such a local framework for prediction can be achieved more generally in quantum cosmology. Such a framework would truly return the observer to importance.}

\section{Inflation in Quantum Cosmology}
\label{examples}

We now turn from illustrative but artificial toy models to more realistic cosmological models. {\rf In this and the next section} we consider two examples in which the return of the observer is important --- where the probabilities for what we observe are significantly different from the probabilities for what occurs.

The two examples share a common theoretical framework \HP. For the dynamics we assume a  spatially closed cosmological  spacetime metric $g$ coupled to  a number of scalar fields $\vec \phi$. The dynamics is specified by a (Euclidean) action $I[g,\vec \phi]$ consisting of the action for general relativity plus the an action for the scalar fields $\vec \phi$ coupled to the metric $g$. For the state $\Psi$ we assume the no-boundary wave function of the universe (NBWF) \cite{HH83}. This is the natural analog of the notion of `ground state' for closed cosmologies. 

Many of our large scale observations are of properties of our universe's classical history. The rate of the universe's expansion {\rf and the distribution of galaxies in our Hubble volume are examples.} A history of the universe behaves classically when the quantum probability is high that it exhibits correlations in time governed by the Lorentzian Einstein equation and the classical field equations. The NBWF predicts an ensemble of alternative classical histories along with third person probabilities for which history in the ensemble occurs. For a little more on what the NBWF is,  and how it predicts probabilities see Appendix \ref{NBWF-mini}. For much more see \cite{HHH08b}. 

{\rf The first example is concerned with the classical cosmological histories predicted by the NBWF} when the matter consists of a single scalar field moving in a quadratic potential
\be
\label{pot}
V(\phi) =\frac{1}{2}m^2\phi^2
\ee
{\rf with $m^2 \ll 1$ and} zero cosmological constant. As in the rest of this paper we are using  Planck units where $\hbar=8\pi G=c=1$.
Geometry and field are restricted to be homogeneous and isotropic thus defining a minisuperspace model.
Lorentz signatured homogeneous and isotropic spacetime geometries can be described by metrics of the form
\be
\label{homoiso-lor}
ds^2 = -dt^2 +a^2(t) d\Omega^2_3. 
\ee
Here, $d\Omega^2_3$ is the metric on a unit round three-sphere. The time-dependence of the scale factor $a(t)$ describes how this closed universe expands and contracts.  Standard closed FLRW cosmological models have metrics of this form that satisfy the Einstein equation.  The homogeneous field is a function only of time, viz $\phi=\phi(t)$. A  quantum history of this model universe is therefore specified by $(a(t),\phi(t))$. Classical histories are described by a pair $(a(t),\phi(t))$ that obey the Einstein equation and the classical equations of motion for the field. 

{\rf As sketched in Appendix \ref{NBWF-mini} the NBWF in this minisuperspace model predicts a one-parameter ensemble of possible classical histories. These classical histories can be labeled by a parameter $\p0$ that can be roughly thought of as the value of the scalar field from which it starts to roll down to the bottom of the potential \eqref{pot}. By examining any one classical $a(t)$ we can find the number of efolds $N_e$ of slow roll inflation it has. Remarkably all histories in the NBWF classical ensemble turn out to have some inflation at early times \cite{HHH08b}. Inflation and the emergence of a classical universe in the NBWF are therefore profoundly connected \cite{HHH08b,Her13}. }

Does our theory \HP predict a significant probability for an extended period of inflation in the early universe? As stated, this question is ambiguous. It could mean ``Are the third-person probabilities from \HP high for classical histories with an early period of inflation?''  But it could also mean ``Are the first person probabilities high that the classical history we observe has a significant period of early inflation?''  We will display the answers to both questions in our model and find that they are significantly different. 

{\rf To evaluate the first person probabilities for the amount of inflation we can assume that we are rare physical systems in any of the classical histories predicted by the NBWF. This is because the number of Hubble volumes in all histories of the classical ensemble with matter densities below the Planck density is much smaller than  any realistic value of $p_E^{-1}$ \cite{HH09}.} Hence volume weighting \eqref{volwgt} connects the first person probabilities for our observations with the third person probabilities for what occurs.

The third person probabilities $p=p(\p0)$ for the histories predicted by the NBWF are given approximately by \eqref{3p-probs-1}
\be
\label{3p-probs}
p(\p0) \propto  \exp[3\pi/V(\p0)] . 
\ee
It turns out in this model that roughly $N_e \approx (3/2)\p0^2$ starting at $N_e \approx 1$.  The third person NBWF probabilities for classical histories \eqref{3p-probs} thus imply  probabilities $p(N_e)$ for the number of efolds that occur, starting roughly at $N_e\approx 1$,
 \be
\label{probefolds-3rd}
p(N_e) \propto \exp\left(\frac{9\pi}{m^2N_e}\right) \ .
\ee
Third person probabilities are therefore much larger for a small number of efolds than for the minimal number $\sim 60$ required for agreement with observation as illustrated in Fig. \ref{efolds}. 

The situation is much different for first person person probabilities for the amount of inflation in {\it our} history --- the one we observe. Since it is assumed that we are rare, the first person probabilities are the volume weighted third person probabilities as in \eqref{volwgt}. During a period of inflation the volume of the universe expands by a factor $\exp(3N_e)$. The first person probabilities are then approximately
\be
\label{probefolds-1st}
\pfp(N_e) \propto \exp\left(3N_e+\frac{9\pi}{m^2N_e}\right) \ .
\ee
Thus the NBWF predicts a significant probability for us {\rf to find ourselves} in a universe with a large number of efolds (Fig \ref{efolds}). This implies there is a significant probability for us to observe the consequences of inflation --- approximate spatial flatness, a scale invariant  spectrum of density fluctuations, etc. 

There is a higher probability for us to live in a larger universe because we are a physical subsystem {\rf within the universe} with a very small probability $p_E$ to have evolved in any Hubble volume. In the larger universes that result {\rf from an extended period of early inflation} there are more Hubble volumes for us to be. Thus the observer returns to importance in the probabilities for observing the consequences of inflation.

\begin{figure}[t]
\includegraphics[width=6in]{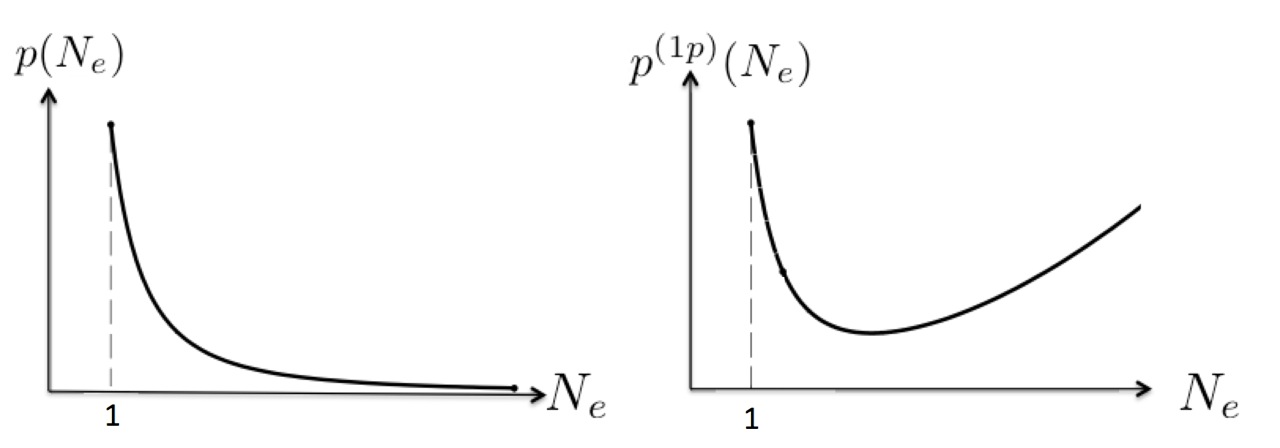}
\caption{\rf Third (left) and first person (right) probabilities for the number of efolds $N_e$ of matter driven inflation in the early universe in the no-boundary quantum state $\Psi$. The first person probabilities favor universes with a large amount of inflation because in the larger universes that result from an extended period of early inflation there are more Hubble volumes for us to be.}
\label{efolds}
\end{figure}

\section{Eternal Inflation and Anthropic Selection}
\label{EI}

{\rf We now extend the above model in three crucial ways to arrive, at last, at a realistic model for the early universe:

(1) {\it Fluctuations:}  We include fluctuations away from homogeneity and isotropy. These fluctuations provide the necessary degrees of freedom to describe e.g. the pattern of temperature variations in the CMB.} A consequence is that a range of classical histories exhibit eternal inflation becoming spatially very large and highly inhomogeneous on the largest scales.  

(2) {\it Landscape Potential:}  We no longer assume one scalar field in a potential with a single minimum like \eqref{pot}. Rather we assume many scalar fields in a multi-field potential that has many different minima with different directions of approach, {\rf as a toy model for the string landscape \cite{Susskind:2003kw}. As a consequence the dynamical theory provides} a mechanism for the observable parameters of the histories to vary.
Automatic anthropic selection can then be explicitly illustrated.

(3) {\it Not Rare but Common:}  The above assumption that we are rare in all histories is no longer tenable with eternal inflation. The probability $p_E$ for us to exist in any Hubble volume is very small. Nevertheless, we will be common in a range of histories where the universe becomes sufficiently large. In that range the connection between third and first person probabilities will no longer involve volume weighting but instead be given by \eqref{common}. 

We now consider the meaning and implications of these extensions in more detail.

\subsubsection{Eternal Inflation -- Histories where Observers are Common}

{\rf Cosmological} perturbation theory extends the homogeneous and isotropic models of Section \ref{examples} straightforwardly to include linear fluctuations away from these symmetries for each classical history. {\rf The early period of slow roll inflation in the classical NBWF histories stretches and amplifies quantum vacuum fluctuations and generates a pattern of classical perturbations on scales larger than the horizon, which much later produce the small temperature fluctuations we observe in the microwave sky. However it turns out that very long-wavelength fluctuations, those which leave the horizon and become classical at values of $\phi$ where 
\be
\label{EIcond}
V^3 \ge |dV/d\phi|^2 \ ,
\ee
have a large expected amplitude.}
There is non-perturbative evidence that histories in which $\phi$ is initially in this regime reach very large (or even infinite) spatial volumes, because these large very long-wavelength fluctuations tend to make them highly inhomogeneous on scales much larger than our Hubble volume \cite{Star86,Linde96,CREM}. {\rf This dynamical process is known as eternal inflation \cite{EIrefs}, and \eqref{EIcond} defines the regime of field values where eternal inflation occurs.}

For quadratic potentials like \eqref{pot}, the condition for eternal inflation \eqref{EIcond} is met for sufficiently large $\p0>\phi_{\rm ei}$ where 
\be
\label{threshei}
\phi_{\rm ei} \sim  1/\sqrt{m} ,
\ee
{\rf well below the Planck scale $\phi_{\rm pl} \sim 1/m$ {\hf for realistic values $m\sim 10^{-5}$. }

{\rf Therefore when we include fluctuations} the set of classical NBWF histories divides into two parts: Those with $\p0\gtrsim\phiei$ which are {\rf inhomogeneous on the largest scales} and have a great many Hubble volumes $N$ due to eternal inflation, and those with $\p0\lesssim\phiei$ {\rf which are much smaller and approximately homogeneous.}

This division has significant implications for the first person probabilities for observation. We are now effectively in the box model case \eqref{both} where (assuming typicality)
\begin{subequations}
\label{both2}
\begin{align} 
[1-(1-p_E)^{N}] \approx 1,   \quad &\p0>\phi_{\rm ei} ,  \label{td1} \\
[1-(1-p_E)^{N}] \approx p_E N \ll 1,  \quad &\p0<\phi_{\rm ei} . \label{td2}
\end{align}
\end{subequations} 
Eternally inflating histories are thus strongly selected, {\rf whereas histories with slow roll inflation only} are strongly suppressed by the very small value of $p_E$.  In selecting for eternally inflating universes as the ones we observe,  the observer has returned in force.

\subsubsection{Landscapes --- A Mechanism for Cosmological Parameters to Vary}
\label{landscapes}

{\mf As discussed in Section \ref{anthropic},  the anthropic selection of observed cosmological parameters requires a  third person theory \HP  that allows the parameters to vary. Theories with landscape potentials are a very simple example. 

To illustrate what landscape potentials are,    consider a third person theory with two scalar fields $\phi_1$ and $\phi_2$ moving in a potential $V(\phi_1,\phi_2)$.  A three dimensional plot of this  potential could be made using $\phi_1$ and $\phi_2$ as the $x-$ and $y-$axes and plotting $V$ along the $z-$axis. The plot might resemble a mountainous landscape on Earth whence the name `landscape potential'. 

Suppose the potential has a number of different minima each surrounded by  a number of different valleys leading to it\footnote{\mf Valleys were called `channels' in \cite{HHH10b} and other places.}. In our past history the fields `rolled down' a particular valley (`our' valley') to a particular minimum (`our' minimum). The value of the potential at our minimum is the value of the cosmological constant we would observe. The shape of our valley near our minimum   determines the spectrum of density fluctuations in the CMB we would see.  A third person theory that predicts probabilities for which of the possible histories occurs is thus a starting point for calculating the first person probabilities for the values of these parameters we observe. 


The notion of a landscape potential is easily extended  to many scalar fields $\vec\phi$ so that $V=V(\vec\phi)$.
Assume that each minimum  in $V(\vec\phi)$ is surrounded by  effectively one-dimensional valleys. 
Suppose that these valleys are separated by large barriers so that transitions between valleys are negligible. Then we have, in effect, an ensemble of one dimensional potentials $V_K(\phi_K)$  $K=1,2,\cdots$ --- one for each valley. }

The NBWF predicts the probabilities for classical inflationary histories for each one-dimensional potential in this landscape. The total classical ensemble is the union of all these. An individual history can therefore be labeled by $(K, \phi_{0K})$ where, as explained before, $\phi_{0K}$ is roughly the value of the field $\phi_K$ at the start of its roll down the potential $V_K$. Thus the NBWF predicts third person probabilities  $p(K, \phi_{0K})$ that our universe rolled down the potential from $\phi_{0K}$ in the valley $K$ to its minimum. 

Generically a landscape of the kind under discussion contains some valleys $K$ where the potential $V_K(\phi_K)$ has a regime of eternal inflation. Eq. \eqref{both2} implies that histories in these valleys will dominate the {\rf first person} probabilities. In the presence of eternal inflation the first person probabilities are
\begin{align}
\pfp(K,\phi_{K0}) &\approx  p(K,\phi_{K0})  \approx \exp[3\pi/V_K(\phi_{K0})], \quad \phi_{k0}\gtrsim \phi_{K\rm{ei}} \nonumber \\
&\approx 0, \quad  \phi_{k0}\lesssim \phi_{K\rm{ei}}
\label{pKp0}
\end{align}
When the potentials are increasing with field, the most probable NBWF history in a given valley will be that where the field starts around the exit of eternal inflation, i.e. $\phi_{K0} = \phi_{K\rm{ei}}$. Thus, to good approximation, the probability that we rolled down in valley $K$ is
\be
\pfp(K) \approx  p(K)  \approx \exp[3\pi/V_K(\phi_{K\rm{ei}})] .
\label{pK}
\ee
We now apply this result in a concrete model landscape.

\subsubsection{First person predictions of the No-Boundary State in a landscape model}
\label{anth-L}

{\rf We now discuss an example of a first person prediction in a landscape model with valleys where the eternal inflation condition \eqref{EIcond} is satisfied.} This example relates directly to the historical effort described briefly in Section \ref{anthropic} to determine the anthropically allowed ranges of cosmological parameters that are consistent with our existence as observers (see, e.g.\cite{Barrow86,Wei89,Tegetal06,LR05}). 

Specifically we calculate the NBWF predictions for the first-person probability of a correlation between three observed numbers: First is the observed value of the cosmological constant $\Lambda \sim 10^{-123}$  (in Planck units). Second is the observed value of the amplitude of primordial density fluctuations $Q\sim 10^{-5}$ in the CMB. And third is the part of our data $\Dl$ on the scales of our galaxy and nearby ones which, together with $\Lambda$, determine the present age of the universe $t_0(\Lambda, \Dloc) \sim 14 {\rm Gyr}$. Thus we are interested in $\pfp(\Lambda, Q, \Dl)$ --- the first person probability for this correlation to be observed. Many other such correlations could be investigated to test a given theory\footnote{For a recent discussion of observables associated with features of the CMB fluctuations see e.g. \cite{HH13}.}. But this example will nicely illustrate several implications of the previous discussion.

We consider a landscape potential in which the parameters $\Lambda$ and $Q$ vary so the theory \HP will predict probabilities for their values. Specifically we assume a landscape of different one dimensional valleys with potentials of the form\footnote{For examples of more general landscapes see \cite{Her13}.} 
\be
V_K(\phi) = \Lambda_K +\frac{1}{2} m_K^2 \phi_K^2,   \quad K=1,2,\cdots .
\label{channel}
\ee
A valley in this landscape is therefore specified by the values $(\Lambda,m)$. (From now on we will drop the subscripts $K$ to simplify the notation.)  

{\rf The NBWF predicts an ensemble of inflationary universes in this landscape}. The observed classical history of our universe rolled down one of the valleys  --- ``our valley''. The values of $\Lambda$ and $m$ that characterize our valley can be determined by observation. Measurement of the expansion history of the universe determines $\Lambda$ and CMB measurements determine $m$, {\rf because the amplitude of the primordial temperature fluctuations is given by \cite{HHH10,HH13}
\be
\label{mQ}
Q \approx N^2_{*}m^2.
 \ee 
Here $N_* \sim{\cal O}( 60)$ is the number of efolds before the end of inflation that the COBE scale left the horizon during inflation.
Observations indicate that $Q\sim 10^{-5} $.}

The first person probability that we observe specific values of $\Lambda$ and $Q$ is the first person probability that our past history rolled down the valley which has those values. 
Since each of the valleys \eqref{channel} has a regime of eternal inflation for $\phi > \phiei \sim 1/\sqrt{m}$ the first person probabilities \eqref{pK} that we observe values of $\Lambda$ and $Q$ can be written as 
\be
\pfp(\Lambda,Q) \approx  p(\Lambda, Q)  \propto  \exp[3\pi/(\Lambda + c \sqrt{Q}/N_*)] .
\label{pLQ}
\ee 
where $c \sim {\cal O}(1)$. For fixed $\Lambda$ this distribution favors small $Q$.
As mentioned above, we are interested in the joint probability $\pfp(\Lambda, Q, \Dl) \approx p(\Lambda, Q, \Dl)$ (first equals third in a regime of eternal inflation). The latter can be rewritten\footnote{In traditional anthropic reasoning the first factor in \eqref{selection-prior}  is called the `selection probability'  which can be calculated;  the second the  is the `prior' which is assumed. {\mf See, e.g. Eq (1) in \cite{Tegetal06}.  The `prior' is typically assumed to be uniform in the anthropically allowed range reflecting ignorance of part of  the theory. }Here both factors follow from the same theory \HP.}
\be
\label{selection-prior}
p(\Lambda, Q, \Dl) = p(\Dl |\Lambda,Q)p(\Lambda,Q).
\ee
The second factor in the expression above is given by \eqref{pLQ}. The first will be proportional to the number of habitable galaxies that have formed in our Hubble volume by the present age $t_0(\Lambda,\Dloc)$. This is plausibly
proportional to the fraction of baryons in the form of galaxies by the present age $t_0$ assuming that is bigger than the collapse time to form a galaxy. Denoting this by $f(\Lambda, Q, t_0)$ we have  from \eqref{pLQ}
\be
\pfp(\Lambda,Q,\Dl) \approx  p(\Lambda, Q,\Dl)  \propto  f(\Lambda, Q, t_0)\exp[3\pi/(\Lambda + c\sqrt{Q}/N_*)] .
\label{pLQD}
\ee 

\begin{figure}[t]
\includegraphics[height=3.5in]{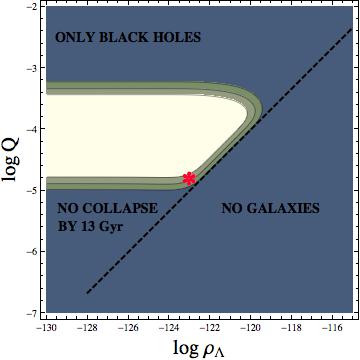}
\caption{A contour plot of the fraction $f(\Lambda, Q, \Dl)$  of baryons in the form of galaxies by the time $t_0 \sim 11 {\rm Gyr}$, adapted from the calculation of  \cite{Tegetal06}. As discussed in the text, the fraction is negligible in the dark (blue) region either because gravitationally bound systems don't collapse or because they collapse to black holes.} 
\label{QLam}
\end{figure}

{All the quantities in \eqref{pLQD} are determinable in one Hubble volume  --- ours. The rest of the calculation can therefore be carried out in that volume. We don't need to ask in which of the vast number of Hubble volumes in the third person eternally inflating universe we dwell. In this model they are all statistically the same as far as the quantities in \eqref{pLQD} go. We are effectively in the situation of the yellow-green box model in Section \ref{cgrain}. We can coarse grain over all other Hubble volumes but ours.  In this way we are able to make contact --- and use --- calculations in traditional anthropic reasoning. }

Figure \ref{QLam} is a contour plot of $f$ at $t_0 \sim 11{\rm Gyr}$ --- earlier than the present age 14 Gyr because galaxies have been around for a while. This was adapted from the detailed astrophysical calculations of \cite{Tegetal06}. For values of $(\Lambda, Q)$ to the right of the diagonal dotted line the universe accelerates too quickly for pre-galactic halos to collapse. Fluctuations have to be large enough to collapse into galaxies a bit before $t_0$ and produce $\Dloc$.  That determines the bottom boundary. If the fluctuations are too large (top boundary) the bound systems are mostly large black holes inconsistent with $\Dloc$. The central region where $f \ge .6$ is the anthropically allowed region. 

The cosmological constant $\Lambda$ is negligible compared to $Q$ in the anthropically allowed (white) region of Figure \ref{QLam} . The exponential dependence $\exp{[3\pi N_*/c\sqrt{Q}]}$ {\rf implied by the NBWF} means that the probabilities \eqref{pLQD} are sharply confined to the smallest allowed values  consistent with {\rf galaxies by $t_0$}, i.e. $Q\sim 10^{-5}$. The resulting marginal distribution for $\Lambda$ is shown in Figure \ref{margL} and is peaked about $\Lambda \sim 10^{-123}$ close to the observed value. 

However the agreement with observation is not the most important conclusion. The model is still too simplified for that. What is important is how the example illustrates the previous discussion. Specifically how anthropic reasoning emerges automatically in quantum cosmology, how it can be {\rf sharpened} by a theory of the universes's quantum state, how first person probabilities select for large eternally inflating universes, and how what is most probable to be observed is not necessarily what is most probable to occur   --- the return of the observer. 

\begin{figure}[t]
\includegraphics[width=3.2in]{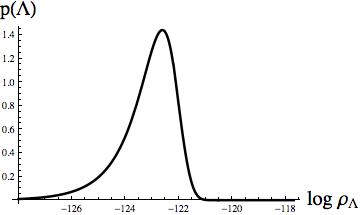}
\caption{The marginal distribution for the cosmological constant $\Lambda$ obtained by integrating \eqref{pLQD} over $Q$.}
\label{margL}
\end{figure}

\section{Conclusion}
\label{conclusion} 

Fundamental physical theories have generally been organized into a third person theory of what occurs and a first person theory of what we observe. In quantum mechanics the output of the two parts are third person probabilities and first person probabilities.  The character of these two parts, their relative importance, and the relation between them has changed over time as new data on new scales of observation required new theory {\rf and as the  understanding of our observational situation in the universe evolved}. The history of the transitions from classical physics, to Copenhagen quantum mechanics, and  then to the quantum mechanics of closed systems that was briefly sketched in the Introduction is part of this evolution.

As we discussed in this paper, modern cosmology  implies new requirements on  both the the first person theory of what occurs  in the universe and the third person theory of what we observe of the universe.  

A  third person theory of the whole universe requires a quantum mechanics of closed systems including quantum spacetime as in \cite{Har95c} for example.
 It requires a quantum mechanics that predicts probabilities for what happens in the early universe when there were no observers around and no measurements being made. It requires a quantum mechanics that predicts probabilities for the emergence of classical spacetime in a quantum theory of gravity.  It requires a quantum mechanics that can explain the origin of the rest of classical predictability in distant realms of the universe that we will never visit.
 
Quantum cosmology  also has implications for first person theory. As observers, individually and collectively, we are physical systems within the universe with only a probability  to have evolved in any one Hubble volume and, a probability to be replicated in many if the universe is very large. What we have seen in this essay is that this implies that  the observer returns to importance in at least the following ways: 
\begin{itemize}
\item By generally showing that what is most probable to be observed is not necessarily the most probable to occur. 

\item By favoring larger universes over small ones if we are rare or if we are common and thus favoring observations of a significant amount of inflation. 

\item  By making anthropic selection automatic. We won't observe what is where we cannot exist.

\item  By requiring the addition of an assumption of typicality to the theory of what occurs made concrete by the xerographic distribution --- an addition that can be tested and used to improve prediction.  

\item By leading to an understanding of how to compute probabilities for our observations of fundamental constants in a landscape that allows them to vary.
\end{itemize}
This list is a brief summary of the results of this essay, but these lead naturally to a number of questions that we discuss here. 

A natural question is why the observer wasn't important in classical cosmology when it is in quantum cosmology. An answer can be traced to differences in starting points and objectives. The starting point for classical cosmology was the assumption that the universe had a single spacetime geometry. The goal was to infer the geometry of that spacetime from large scale observation. Is it approximately homogeneous and isotropic on large scales, open or closed... ? What are the values of the cosmological parameters that characterize it --- the cosmological constant, the Hubble constant, the amount of radiation, the amount of baryons, the amplitude of the density fluctuations, etc? Is there evidence that the spacetime had an early period of inflation? Observers were presumed to exist but had a negligible influence on the answers to these questions.

Quantum cosmology does not start by assuming classical spacetime. Rather it  starts from a theory of the universe's quantum state and dynamics. From that it seeks to explain when spacetime is classical and predict probabilities for what the different possible classical spacetimes are  --- questions that cannot even be asked in the context of classical cosmology. It therefore answers to the classical questions with probabilities about large scale geometry.

A second natural question is  {\it why} are fundamental  physical theories encompass both a third person theory of what occurs and a first person theory of what is observed? The historical success of theories organized in this way is  indisputable. That success tells us something about the world. It supports the idea of some form of realism, perhaps along the lines of what Putnam called `realism with a human face' \cite{Putnam}. To paraphrase J. A. Wheeler `In a quantum world, the universe is a grand synthesis, putting itself together all the time as a whole. Its history {is} ... a totality which includes us and in which what happens now gives reality to what happened then.' \cite{Wheeler}

A third natural question concerns the scale on which we have to know something of the universe to make first person predictions. An IGUS\footnote{Information gathering and utilising system.} like us is confined to a very local region inside of a vast Hubble volume which itself is typically but a small part of a much larger universe. Yet the present formulation of the first person theory requires information beyond our Hubble volume to determine whether we are rare, or common, or other. {\lf It remains to be seen whether a more local computation of probabilities for observation in quantum cosmology can be found.} That would be a further way the observer and universe are unified. 

Everett's insight was that, as observers of the universe, we are physical systems within it, not outside it. We are subject to the laws of quantum mechanics but play no special role in its formulation. We are negligible perturbations on a third person description of the universe. But, as shown in this essay in several different ways , we return to importance for calculating  first person probabilities for our observations precisely because we are physical systems in the universe. We may have  only begun to appreciate the ramifications of that insight.

\noindent{\bf Acknowledgments:} The authors have had the benefit of collaboration and discussions with a great many scientists in this work. The cited papers have those acknowledgments. However, our collaborators on the papers that underly this work should be thanked. They are Murray Gell-Mann, Stephen Hawking, and Mark Srednicki. {\jf Gregory Benford suggested the title.} {\hf Comments of Don Page and Simon Saunders were useful.} {The work of JH was supported in part by the US NSF grant PHY12-05500. The work of TH was supported in part by the National Science Foundation of Belgium (FWO) grant G.001.12 Odysseus and by the European Research Council grant no. ERC-2013-CoG 616732 HoloQosmos. 

\appendix

\section{Typicality --- The Xerographic Distribution}
\label{typicality}

Consider for a moment a  third person theory --- classical or quantum ---that predicts a large number of Hubble volumes, some with one kind of observable property, some with another. A box model like those of Section \ref{boxmodels}  with one history and different colored volumes would be a very simple example.

Suppose that the data $D_0$ that describe our observational situation (including us) occur in many different Hubble volumes. One of them is ours, but in other volumes the result of the observation specified by $D_0$ could be different. A third person theory \HP would predict what is observed in all of the instances of $D_0$. But it does not predict which one is ours. Indeed, it has no notion of `we' or `us'. 

We don't know which of these copies of $D_0$ are us. It could be any one of them. To make {\rf first person} predictions for our observations, the third person theory must therefore be augmented by an assumption about which instance of $D_0$ is us, or more generally with a probability distribution on the set of copies. If there is no such assumption there are no predictions. {\rf Put differently, predictions for our observations require a statement on what exactly we mean by `us'  - a specification of how we think our particular situation relates to other instances of $D_0$ in the universe.}
 
In the body of this essay we have consistently assumed that we are equally likely to be any of the incidences of our data $D_0$ that the third person theory \HP predicts. This is the simplest and least informative assumption but it is not the only possible one. Other, more informative  assumptions  may lead to better agreement between theory and observation, be more justified by fact, and be more predictive\footnote{As for the Boltzmann brain problem \cite{HS07}.  Boltzmann brains are not a problem if we assume that our observations are not typical of deluded observers who only imagine that they have the data $D_0$. }.  

We call a distribution that gives the probability that we are any of the incidences of $D_0$ or a subset of it  a {\it xerographic distribution} \cite{SH10}. It is usually written $\xi_i$ where the index ranges over all incidences of $D_0$. A xerographic distribution is effectively a formal expression of the assumption about how typical we are in the universe in the set of all other incidences of $D_0$. 

It is the theoretical structure consisting of \HP and $\xi$ that yields first person predictions for observation. Each of the elements in this combination is therefore testable by experiment and observation. Just as we can compete different \HP by their predictions for observation we can also compete different $\xi$.

\section{The No-Boundary Wave Function in Homogeneous and Isotropic Minisuperspace}
\label{NBWF-mini}
{\rf The example} in Section \ref{examples} assumed the no-boundary wave function (NBWF) for the quantum state $\Psi$ in the third person theory \HP.  The NBWF is the analog of the ground state for closed cosmologies and therefore a natural candidate for the wave function of our universe. Its predictions for observations are in good agreement with observation. For example, it predicts that fluctuations away from homogeneity and isotropy start out near the big bang in their ground state \cite{Halliwell85}. {\rf Combined with its prediction that our universe underwent an early period of inflation \cite{HHH08a}} that leads to good agreement with the observed fluctuations in the CMB. This Appendix provides the reader a little more explanation of what that wave function is and how its consequences used in the examples are derived. For more detail see \cite{HH83}. 

A quantum state of the universe like the NBWF is represented by a wave function on a configuration space of three geometries and matter field configurations on a spacelike surface $\Sigma$. On the minisuperspace of homogeneous and isotropic geometries \eqref{homoiso-lor} and a single homogeneous scalar field we write 
\be
\label{qs}
\Psi = \Psi(b,\chi).
\ee 
Here $b$ is the scale factor of the homogeneous, isotropic metric \eqref{homoiso-lor} on $\Sigma$  and $\chi$ is the value of the homogeneous scalar field. The no boundary wave function is a particular wave function of this form.

The NBWF is formally defined by a sum over homogeneous and isotropic Euclidean geometries that are regular on a topological four-disk and match $b$ and $\chi$ on its boundary \cite{HH83}, weighted by $\exp(-I)$ where $I$ is the Euclidean action of the configurations.  We won't need to go into this construction. 
We will only need its leading order in $\hbar$ semiclassical approximation. That is given by the saddle points (extrema) of the action $I[a(\tau),\phi(\tau)]$ on this disk for metric coupled to scalar field. There is one dominant saddle point for each $(b,\chi)$. Generally these saddle points will have complex values of both metric and field. The semiclassical NBWF is a sum over saddle points terms of the form (in units where $\hbar=1$)
\be
\label{saddle}  
\Psi(b,\chi)  \propto  \exp[-I(b,\chi)/\hbar] \propto \exp{[-I_R(b,\chi) +iS(b,\chi)] }\                  
\ee
where $I(b,\chi)$ is the action evaluated at the saddle point and $I_R$ and $-S$ are its real and imaginary parts. 

The wave function \eqref{saddle} has a standard WKB semiclassical form. As in non-relativistic quantum mechanics, an ensemble of classical histories is predicted in regions of configuration space where $S$ varies rapidly in comparison with $I_R$.  The histories  are the integral curves of $S$ defined by solving the Hamilton-Jacobi relations relating the momenta $p_b$ and $p_\chi$ conjugate to $a$ and $\phi$ to the gradients of $S$
\be
\label{intlecurves}
p_b =\nabla_b S, \quad\quad  p_\chi=\nabla_\chi S .
\ee

There is a one parameter family of classical histories  --- one for each saddle point.  It is convenient to label them by the magnitude of the scalar field $\phi_0$ at the center of the saddle point. One can think crudely of $\p0$ as the value of the scalar field  at which it starts to roll down to the bottom of the potential in a classical history. 

The third person probabilities for these histories are  proportional to the absolute square of the wave function \eqref{saddle}
\be
\label{3p-probs-1}
p(\p0) \propto \exp[-2I_R(\p0)] \approx  \exp[-3\pi/V(\p0)] . 
\ee
The last term is a crude approximation to the action which is useful in rough estimates \cite{Lyo92}. 

This is a very quick summary of a lot of work. For more details see \cite{HHH08b}. In short, in the leading semiclassical approximation the NBWF predicts and ensemble of possible classical spacetimes obeying \eqref{intlecurves} with third person probabilities \eqref{3p-probs}.

\end{document}